\documentclass{iaus}
\usepackage{graphicx}

\title[~~Far Infrared Luminosity Function in ADF-S] 
{Far Infrared Luminosity Function of Local \\Galaxies in the AKARI Deep Field South}

\author[Chris Sedgwick et al.]   
{Chris Sedgwick$^1$,
Stephen Serjeant$^1$,
Chris Pearson$^{4,2,1}$,\\
Shuji Matsuura$^{3}$,
Mai Shirahata$^{3}$,  
Shinki Oyabu$^{8}$, 
Tomotsugu Goto$^{5,6}$,
Hideo Matsuhara$^{3}$,
D.L. Clements$^{7}$,
Mattia Negrello$^{1}$,\\
Toshinobu Takagi$^{3}$,
\and Glenn J. White$^{1,2}$}

\affiliation{$^{1}$Department of Physical Sciences, The Open University, Milton Keynes MK7 6AA\\
$^{2}$Rutherford Appleton Laboratory, Chilton, Didcot, Oxfordshire OX11 0QX\\
$^{3}$Institute of Space and Astronautical Science, Japan Aerospace Exploration Agency, Sagamihara, Kanagawa, 252 5210, Japan\\
$^{4}$Institute for Space Imaging Science, U. of Lethbridge, Lethbridge, Alberta, T1K 3M4, Canada\\
$^{5}$Institute for Astronomy, U. of Hawaii, 2680 Woodlawn Drive, Honolulu, HI 96822, USA\\
$^{6}$Subaru Telescope, 650 North A'ohoku Place, Hilo, HI 96720, USA\\
$^{7}$Astrophysics Group, Imperial College, Blackett Lab., Prince Consort Rd, London SW7 2AZ\\
$^{8}$Graduate School of Science, Nagoya U., Furo-cho, Chikusa-ku, Nagoya, Aichi 464-8602, Japan}

\pubyear{2011} 
\volume{284}  
\pagerange{1--12}
\jname{The Spectral Energy Distribution of Galaxies}
\editors{R.J. Tuffs \&  C.C.Popescu, eds.}
\begin{document}

\maketitle

\begin{abstract}
We present the first far-infrared luminosity function in the AKARI Deep Field South, a premier deep field of the AKARI Space Telescope, using spectroscopic redshifts obtained with AAOmega. To date, we have found spectroscopic redshifts for 389 galaxies in this field and have measured the local ($z<0.25$) 90 $\mu$m luminosity function using about one-third of these redshifts. The results are in reasonable agreement with recent theoretical predictions. 
\end{abstract}

\firstsection 
              
\section{AKARI Deep Field South}

The AKARI Deep Field South (ADF-S) is centered on RA 4h 44m 00s, Dec -53$^o$ 20\textsf{'} 00\textsf{"} (J2000), and is a low-cirrus (I$_{100\mu m}$$<0.5$ MJy sr$^{-1}$) region of about 12 square degrees near the South Ecliptic Pole (Matsuhara et al. 2006). It has been extensively studied over the full area by the AKARI-FIS (at 60, 90, 140, and 160 $\mu$m, Shirahata et al. 2008), Spitzer (24 and 70 $\mu$m, Clements et al. 2011), BLAST (250, 350, 500 $\mu$m, Valiante et al. 2010), and over part of the area by CTIO (UVBRI), AKARI-IRC (7 mid-infrared bands, Pearson et al. in preparation), ATCA (deep radio survey at 1.4 GHz, White et al. 2011), ASTE/AzTEC (1.1 mm, Hatsukade et al. 2011) and APEX/LABOCA (870 $\mu$m, Khan et al. in preparation). Other work includes CIB detection (Matsuura et al. 2011) and cross-identification with public data and SEDs (Malek et al. 2010). A large part of the field is being mapped by Herschel as part of the HerMES project.

The 90 $\mu$m catalogue from the AKARI Far Infrared Surveyor (FIS) deep survey lists 2,282 sources at a signal-to-noise ratio SNR $>5$ (giving a minimum flux of 12.81 mJy). Since this wavelength is within the range at which the re-processed radiation received from dust in Star Forming Galaxies (SFGs) is expected to peak, the evolution of the 90 $\mu$m luminosity function provides an excellent proxy for the study of star formation. 

This poster (which is based on Sedgwick et al. 2011) describes the first luminosity function from this data, for $z<0.25$ galaxies, and combines this data with earlier ISO data from the ELAIS survey to provide the highest signal-to-noise deep luminosity function at this wavelength to date. We also compare our results with predictions from a recent backward galaxy evolution model, finding a reasonable agreement. Our aim is to provide a local benchmark against which to compare luminosity functions of more distant SFGs based on more detailed, deeper data which will become available in the near future.


\section{AAOmega Spectroscopy}

We have used AAOmega to obtain spectra of selected infrared and sub-millimetre sources with optical counterparts in 3.14 square degrees in the centre of the ADF-S. Data was obtained between October 2007 and November 2008 and redshifts identified for 389 sources, of which we have used 130 for our luminosity function.

\begin{figure*}
{\includegraphics[width=2.6in]{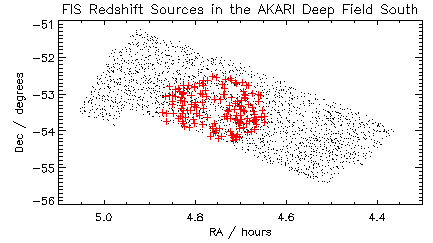}}
{\includegraphics[width=2.6in]{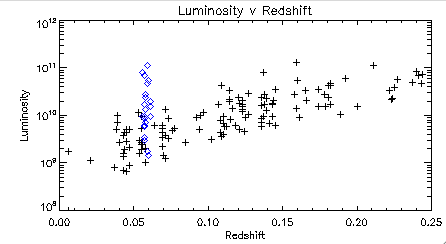}}
\caption{Left: FIS sources with z$<$0.25 (red plus symbols) on all FIS sources (black dots) in the AKARI Deep Field South. Right: the redshift - luminosity plane for all sources used, with non-cluster sources shown as black plus symbols and the cluster sources shown as blue diamonds. Luminosity is $\nu$L$_\nu$/L$_{\odot}$ as calculated for the luminosity functions in this paper. }\label{fig:field}
\end{figure*}

Redshifts were measured using the [NII]/H$\alpha$/[NII], and H$\beta$/ [OIII]/[OIII] emission lines. Redshifts were only taken where one or other (or both) of these two triplets of emission lines were identified. Unless out of range, the [SII] doublet was also seen. The [OII] emission line resulting from the $\lambda\lambda$3726, 3729 was often found at the same redshift to support the identifications. AGNs and SFGs were segregated using a BPT diagram. Figure \ref{fig:field} (left) shows the redshifts $z<0.25$ which were used for this paper.

As shown in the luminosity/redshift plot (Figure \ref{fig:field} right), we have identified a new cluster in this field at $z \sim 0.06$, centred on RA 4h 42m 12.5s, Dec -53$^o$ 30\textsf{'} 46\textsf{"} which is discussed in Sedgwick et al. (2011) and excluded from the luminosity function.

\section{Far-infrared Local Luminosity Function}

In the absence of completeness as a function of flux based directly on AKARI data (under construction, Shirahata et al.), we have used the ultra-deep Spitzer 70 $\mu$m observations of the GOODS-N field which reached down to a flux of 1.2 mJy. Frayer et al. (2006) showed that this is well fitted by the model of Lagache et al. (2004), which we used to generate number counts against which to measure AKARI completeness. We have adjusted for our 90 $\mu$m case using the model described in Pearson (2001), scaling the fluxes by a factor of 0.673 on the basis of its prediction that N(S$_{70\mu m}$ $>0.0673$ Jy) = N(S$_{90\mu m}$ $ >0.1$ Jy). In addition, we have estimated spectroscopic completeness by comparing the magnitude distribution of AAOmega redshift sources with that of APM B-magnitude sources which have counterparts in the total ADFS-FIS catalogue. The area of the sample was re-normalised to account for large scale cosmic variance, using the number density of objects $>0.1$ Jy in the 50 deg$^2$ SWIRE survey (following the procedure in ELAIS,  Serjeant et al. 2001). We calculated the 1/V{\tiny max} luminosity function (Schmidt 1968) following the methodology in Serjeant et al. (2004). K-corrections were made assuming the M82 star-forming spectral energy distribution. We have assumed cosmological parameter values of H$_0$=72 kms$^{-1}$Mpc$^{-1}$, $\Omega$$_M$=0.3 and $\Omega$$_{\Lambda}$=0.7.

\begin{figure*}
 {\includegraphics[width=1.75in]{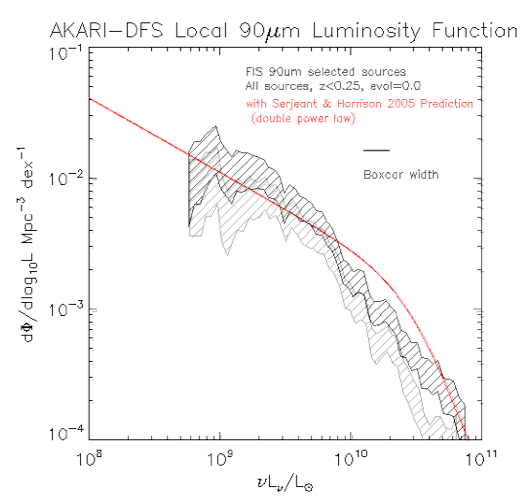}}
 {\includegraphics[width=1.75in]{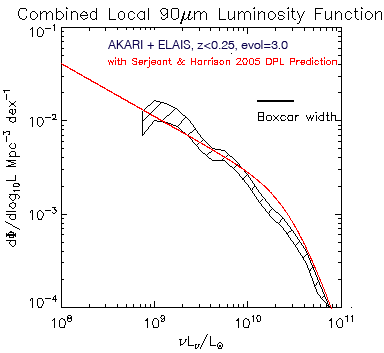}} 
 {\includegraphics[width=1.65in]{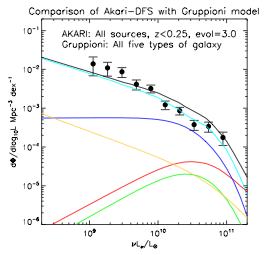}}
\caption{Left: Our new ADF-S luminosity function (Poisson error bands hatched; SFGs-only in shadow). The line is the double power-law prediction by Serjeant \& Harrison 2005. Centre: Combined AKARI and ELAIS luminosity function. Right: Comparison of our (unsmoothed) data with predictions from the Gruppioni et al. (2011) model. Lines (from top at 10$^{10}$): total (black), spirals (cyan), Seyfert2s (blue), AGN1s (brown), starbursts (red), obscured AGNs (green).}\label{fig:combined_lf}
\end{figure*}


In Figure \ref{fig:combined_lf}, the left panel shows our luminosity function assuming pure luminosity evolution of (1+z)$^3$. We have combined our results with the 90 $\mu$m luminosity function of the ELAIS-N fields presented in Serjeant et al. (2004) in the centre panel (as described in detail in Sedgwick et al. 2011),  and in the right panel we show that our results are in reasonable agreement with the theoretical prediction for the 90 $\mu$m luminosity function for $0<z<0.25$ galaxies from a new backward evolution model of Gruppioni et al. (2011) which includes separate evolution for five types of galaxy. The figure also shows that our result is close to the Serjeant \& Harrison (2005) double power-law prediction, with a slight faint-end excess and a deficit just below L*, attributed to large-scale structure.


{\underline{\it Acknowledgements.}} This research is based on observations with AKARI, a JAXA project with the participation of ESA. This work was funded in part by STFC (grant PP/D002400/1), the Royal Society (2006/R4-IJP), the Sasakawa Foundation (3108) and KAKENHI (19540250 and 21111004). We extend our thanks to Carlotta Gruppioni.

\end{document}